\documentclass[twocolumn]{revtex4}

\def\be{\begin{equation}}
\def\ee{\end{equation}}
\def\la{\label}
\def\bea{\begin{eqnarray}}
\def\eea{\end{eqnarray}}

\def\ci{\cite}
\def\la{\label}
\def\bib{\bibitem}

\def\Lm{\Lambda}
\def\Lmc{\Lambda_c}
\def\le{\left}
\def\ri{\right}

\def\Ompi{\Omega_{\phi i}}

\def\Om{\Omega}
\def\Omq{\Omega_{DG }}
\def\Omqi{\Omega_{DG i}}
\def\Omqo{\Omega_{DG o}}
\def\Omde{\Omega_{DE}}

\def\s8{\sigma_8}

\def\fr{\frac}
\def\pp{\partial}

\usepackage{graphicx}

\begin{document}

\title{A Realistic Particle Physics Dark Energy Model}

\author{A. de la Macorra}
\affiliation{Instituto de F\'{\i}sica, UNAM, Apdo. Postal 20-364,
01000 M\'exico D.F., M\'exico}

\begin{abstract}

We present a realistic dark energy model derived from particle
physics. Our model has essentially no free parameters and has an
equivalent fit to the observational data (CMB, SN1a and LSS)   as
LCDM and a better fit than the best effective $w(z)$ model.
With the lack of a clear determination of the cosmological
parameters theoretical considerations should be taken seriously to
distinguish between dark energy models.

\end{abstract}


\maketitle

Recent cosmological observations \ci{wmap}-\ci{LSS} imply
the universe is dominated by a dark energy fluid. However, there is no theoretical
understanding what this dark energy is.  The simplest
and most common assumption is
a cosmological constant model. Comparing
 LCDM to wmap, SN1a and LSS data one finds
a large  degeneracy among the
cosmological parameters
$P_6=(H_o,\Omde,\Omega_bh^2,\tau,n_s,A_s)$
 \ci{tegmark}-\ci{seljak}.
Several attempts have been made to infer from  the data
the best dark energy model by parameterizing the equation
of state parameter $w(z)$ as a function of redshift $z$ \ci{wz}.
  The analysis shows that
 the degeneracy of the cosmological parameters $P_6$
remains but the central values   varies. The best fit model
for a dynamical $w(z)$ and LCDM have an  equivalent
 $\chi^2/dof=1.08$ value \ci{tegmark},\ci{wz} but LCDM has a better
 Akaike  or Bayesian   information criteria \ci{wz}.
With the lack of a clear determination of $P_6$
theoretical considerations should be taken seriously to
distinguish between dark energy models.

Even though the LCDM  model  has a good fit to the cosmological data it is
hard to accept it from a theoretical point of view. It does not
answer some fundamental questions as where it comes from and why
does it have at present time an energy of the same order of
magnitude as dark matter, i.e. the coincidence problem. From a theoretical point
of view we would like to have a dark energy model which is well
motivated from particle physics, answers the fundamental question
which a LCDM  does not  and has at least an equivalent fit to the
data. The best theoretical model should have the least number of
free parameters and should give  reasonable explanations and values for them.

In this work we will present a realistic dark energy model derived
from particle physics. We will show that this model has, indeed,
an equivalent fit to the observational data as LCDM. It has
essentially  no free parameters so the cosmological $P_6$ quantities
are the only free parameters. For most of the cosmological time the energy
density is proportional to radiation since the particles are massless.
It is the dynamics of the
model that give rise to a late time phase transition at a scale
$\Lm_c=O(10-100 eV)$ and an effective scalar field appears which
gives the dark energy today. For energies larger than $\Lm_c$ the
particles are massless and redshift as radiation and below $\Lmc$
the particles  acquire a non-perturbative mass. The model is based on
gauge theory and has an equivalent structure as the well
established standard model of particle physics. The dark energy gauge group
gives an effective potential of the
form $V\propto \phi^{-n}$, where $\phi$ is the dark energy
(quintessence field) and the power $n $ is fixed by the number of fields.
Inverse power low potentials "IPL" were first introduced by
\ci{ratra} and  \ci{bine}  to use ADS superpotentials
\ci{ADS}
to get an effective IPL potential.
However, the models of \ci{ratra} had no particle physics
interpretation and the models of \ci{bine}  had $n>2$
and are phenomenological not viable. We were the first ones to
study phenomenological acceptable ADS superpotentials giving an
IPL with $n<2$ and  our model has $n=2/3$ \ci{ax.1}.
We would like to emphasis the following theoretical points of our model:\\
$\bullet$ The potential is calculated from  our best theoretical
theories, i.e. gauge theories \\
$\bullet$ The superpotential is exact  and the potential is stable against quantum corrections\\
$\bullet$  The appearance of quintessence field is  due to   a late time
  phase transition given by  $\Lmc$ \\
$\bullet$  The condensation scale $\Lmc$ is determined by first principles
(it is given in terms of  the number of fields)\\
$\bullet$  The initial energy  is given also only  by the number of fields\\
$\bullet$  The number of fields   is determined by imposing  gauge coupling unification  \\
$\bullet$  The solution is an attractor even though the quintessence field has not
reached it yet\\
$\bullet$ Our dark energy model has no free parameters\\
$\bullet$ It has an equivalent fit to the observational data (CMB and SN1a)  as the best fit model\\
$\bullet$ It "solves" the
coincidences problem\\
This last point is due to a late time phase transition. The onset of the quintessence
field is at about matter-radiation equality since the scale factor $a_c$
(i.e. at  $\Lmc$)
is in cosmological times very close to matter-radiation equality
$a_{eq}$ ( $a_c/a_{eq}\simeq 10^{-2}$). This means that
$\rho_{DG}/\rho_{r}$ is constant from the end of inflation $a_{inf}\simeq
10^{-30}$ until $a_c \simeq
10^{-6}$ which accounts for most of the time.

The dark energy model is simply a $SU(N_c=3)$ gauge group with
$N_f=6$ elementary particles in the fundamental representation and
with only gravitational interaction with the standard model of
particle physics. The phase transition scale $\Lmc$ is determined
only by $N_c$ and $N_f$ and the value of the
 gauge coupling constant at some arbitrary
energy scale. Motivated by string theory our dark energy model is
constrained to be unified with the standard model gauge groups at
the unification scale. For viable cosmological models
gauge coupling unification  determines the values of $N_c,N_f$
\ci{ax.1}.
With this non-trivial constrained the evolution of the gauge
coupling constant    and $\Lmc$ are
completely fixed.   At high
energies the dark  elementary fields are massless and
$\rho_{DG}\propto a^{-4}$, $\rho_{DG}/\rho_r$ is constant and the ratio
is given only in terms of the number of particles (c.f. eq.(\ref{TNS})).
At lower energies a
phase transition takes place due to a strong gauge coupling
constant. At this scale the dark elementary fields are bind
together producing gauge invariant states. The relevant scale for
this process is the condensation scale $\Lm_c$ and for  a  gauge
group $SU(N_c)$ with $N_f$   matter fields in susy is given
by the one-loop renormalization group equation $\Lm_c= \Lm_{gut}
e^{-8\pi^2/b_o g^2_{gut}} $ where $b_o=3 N_c-N_f$ is the one-loop
beta function and $\Lm_{gut} \simeq 10^{16}\mathrm{GeV},
g_{gut}\simeq \sqrt{4 \pi/ 25.7}$   are the unification
energy scale and coupling constant, respectively. In our  dark
energy model we have $N_c=3,N_f=6$ giving $b_o=3$ and
 $\Lm_c=42\,eV$. Strong gauge interactions produce a non-perturbative
potential $V$. This potential can be calculated using
ADS potential \ci{ADS}. The superpotential for a non-abelian
$SU(N_c)$ gauge group with $N_f$   massless fields  is
 $ W=(N_c-N_f)(\Lm_c^{b_o}/det \langle Q\tilde
Q\rangle)^{1/(N_c-N_f)}$. The scalar potential  in  susy
for one dynamical meson
field $\phi$, the pseudo-Goldstone boson, is $V=e^{\phi^2/2}|W_\phi|^2$ with $W_\phi=\pp
W/\pp \phi$,  giving
 \ci{bine},\ci{ax.1} \be
V=c^2\Lm_c^{4+2/3}\;e^{\phi^2/2}\;\phi^{-2/3}
 \la{vsugra} \ee
with $c=2N_f=12$. We have included the exponential factor in
eq.({\ref{vsugra}) because once gravity is taken into account it gives
the correct normalization in the Einstein frame. For
values of the field much smaller than the Planck mass $m_p=2.4\,10^{18}\,GeV \equiv 1$
the exponent term is irrelevant but for $\phi\simeq 1$ it gives a
(small) correction. The evolution is basically determined
by the inverse power law exponent $\phi^{-2/3}$.
 In order to
study the evolution of eq.(\ref{vsugra}) the initial condition on
$\phi$ must be set and the only natural initial value  is
$\phi_i=\Lm_c$ since it is precisely $\Lm_c$ the relevant scale of
the physical binding process. The potential in eq.(\ref{vsugra})
has a minimum at $\phi=\sqrt{n}= \sqrt{2/3}$ and the value of the
potential   is $V|_{min}=c^2\Lmc^{4+2/3}e^{1/3}(3/2)^{1/3}$. The
evolution of $\phi$ with different initial conditions is shown in
fig.{\ref{grphi}. Even though the evolution of $\phi$ depends on
$\Ompi$ its value at present time $a_o=1$ is (almost) independent
of the initial conditions. Global symmetries and susy protect the
mass of the quintessence field $\phi$. In fact the ADS
superpotential is exact and receives no corrections \ci{ADS}. The
only radiative corrections arise due to the Kahler potential.
However, our model is stable against radiative corrections
\ci{ax.mod}, the shape of the potential remains the same and the
v.e.v. of $\phi$ suffers a small shift but this  does not affect
the cosmological predictions since the solution is an attractor.
\begin{figure}[tbp]
    \includegraphics[width=8.5cm]{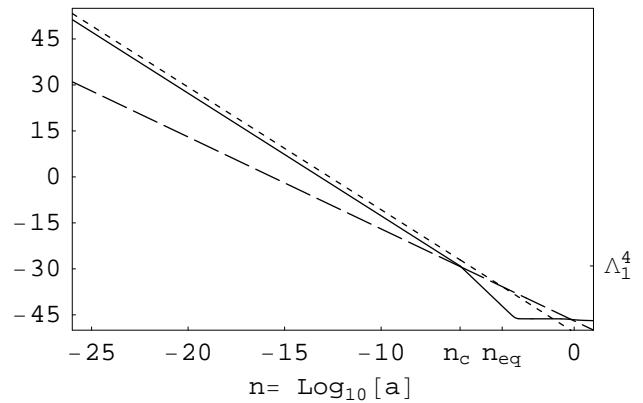}
 \caption{We show the evolution of $\rho_r,\rho_m$ and $\rho_{DG}$ (dotted, dashed and solid lines,
  respectively) as a function of $Log_{10}(a)$
  with $a_o=1$ with  no extra degrees of freedom, i.e.
$g_h=0$. The value of equivalence is $n_{eq}=-3.6$ while
the phase transition takes
  place at $n_c=-5.9$.  }
 \label{grrho}
\end{figure}
\begin{figure}[tbp]
   \includegraphics[width=8.5cm]{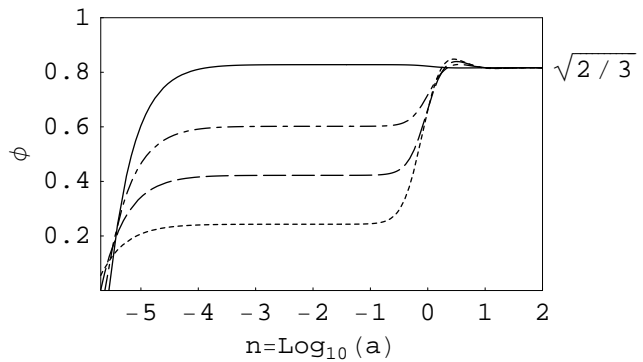}
  \caption{ We show the evolution of $\phi$ for different
  initial conditions $\Omqi=0.01,0.03,0.06,0.11$  (dotted, dashed,
   dot-dashed and solid    lines, respectively).}
 \label{grphi}
\end{figure}
The complete evolution of the dark energy group can be seen in
fig.\ref{grrho}. Notice that $\rho_{DG}$ tracks radiation for a
long period of time, including nucleosynthesis "NS" epoch, since
all the particles are massless and the onset of the quintessence
field is at a very late time ($a_c\simeq 10^{-6}$) and close to
matter-radiation equality ($a_{eq}\simeq 10^{-4}$). The fact that
$\Lmc$ is so small and the appearance of the quintessence field is
at such a late time "solves" the coincidence problem since it
implies that an accelerating universe will necessarily be at a
scale factor larger than  $a_c $ and $a_{eq}$, close to present
time. In our case the scalar field is present only at $a\geq a_c$
and it does not track the background at all, as
  can be seen from figs.\ref{grrho},\ref{grsugraw} and \ref{grsugrawz}.

The total energy density of the universe is given
$\rho_T=\rho_{sm}+\rho_{DG} +\rho_{h}$, where $\rho_{sm}$ is
the energy density of the standard model (we will take it as
the minimal supersymmetric standard model) and we have allowed for
$\rho_{h}$ hidden particles which could be present at earlier
times. Examples of these particles are the ones requires to break
susy in the standard model and are widely predicted by string
theory. The only constrained we imposed to $\rho_{h}$ is that they do
not contribute to the present day energy density nor at NS.
One important constrained on the energy density at  earlier times
is given at NS. The extra amount of
relativistic energy density at NS is usually parameterized by an effective
neutrino $\Delta N_\nu\leq 0.2-0.3$ which implies an upper bound
$\Om_{ex}(NS)\leq 0.045$ in the most stringent case
and $\Om_{ex}(NS)\leq 0.09$ at two $\sigma$ \ci{NS}. However,    NS is  still
fine with $\Delta N_\nu\leq 4.1 $ which implies a bound
$\Om_{ex}(NS)\leq 0.4$ if new physics is involved at that time (as
neutrino asymmetry) \ci{NS2}.

Our dark energy model has no free parameter, however,  the total
energy density could in principle  contained more particles than
the standard model and our dark group, i.e. $g_h \neq 0$.
In fact extra particles would be needed to break susy in the
visible sector. To simplify the discussion we will assume that
these particles, if present, acquire a large mass (larger than
$TeV$) so that they do not affect NS   and that they are coupled
to the standard model at high energies. Examples of this are the
gauge mediated susy breaking models, extensively studied in
particle physics context, which require a energy scale  $\Lm_h=O(10^{7-8}) GeV$ in
order to give masses of the order of $TeV$ to the standard model particles \ci{susygauge}.
   Since the hidden sector is only coupled to our dark
model via gravity, it will induce a susy breaking to the dark
group  of the order
 $m_{s}=\Lm_{h}^3/m_p^2 $ which is smaller than $\Lmc$ and
 does, therefore,
 not affect  the
 running of the gauge coupling and the onset of our potential
 given in eq.(\ref{vsugra}).
These extra particles will  not contribute to the
energy density at NS nor at present time but they are important
for determining the ratio of temperature between the standard
model and our dark group.
The effect of these extra
particles is to decrease  the ratio of the temperature and
$\Omq$. From entropy conservations one can determine $\Omq$
giving \ci{ax.mod}
 \be\la{TNS}
\Omq=\fr{g_{Q }(T_D/T)^4}{g_{sm}
 +g_{Q }(T_D/T)^4},\hspace{.5cm} \le(\fr{T_D}{T}\ri)=\le(\fr{q}{g_{dec}}\ri)^{1/3}
 \ee
 with $q=10.75,\,g_{sm}=10.75$ at NS and $q=2\times10.75\times4/11,\,
 g_{sm}=3.36$ at $\Lmc$ and
 $g_{dec}=g_{MSSM}+g_{h}$, with $g_{MSSM}=228.75$  and $g_{h}$ the degrees of freedom
 of the MSSM and the hidden sector, respectively. The term 4/11
 in $q$ at $\Lmc$
 takes into account for neutrino decoupling and $10.75, 3.36$ are the relativistic
 degrees of freedom of
 the standard model at NS and at present time while $g_Q=97.5$ are the degrees
 of freedom of our model
 \ci{ax.mod}.  An
upper bound on $\Omq, T_D$ can be established if there are
no hidden particles,  $g_{h}=0$, and   eq.(\ref{TNS}) gives
 \be \Omq(NS)\leq 0.13 \hspace{1cm}
\Omqi(\Lmc)\leq 0.11
 \ee
where  $\Omqi$ stands at $\Lmc$, i.e. the onset of the
quintessence field.
Notice that  the upper limit is still
within the existing NS bounds (not
the most stringent one)   since  $\Delta N_\nu \leq 1$ implies
$\Omega_{ex}\leq 0.14$. For $g_{h}\neq 0$ than we get a
smaller $\Omq(NS)$ and for $\Omq(NS)=0.09,\,0.045$ we
require $g_{h}=90, \,327$, respectively, giving
$\Omqi=0.076,\,0.037$. Typical gauge mediated susy
breaking models have $g_{h}=O(200)$ \ci{susygauge}.
Clearly  we have no precise knowledge of how many hidden sector
particles are nor how susy is broken. Furthermore,
the contribution of $g_{h}$ would be the same for all  dark
energy models and the
uncertainty  lies in the hidden  sector and not on our dark group
model   which, as stated above,  has no free parameter.
\begin{figure}[tbp]
    \includegraphics[width=8.5cm] {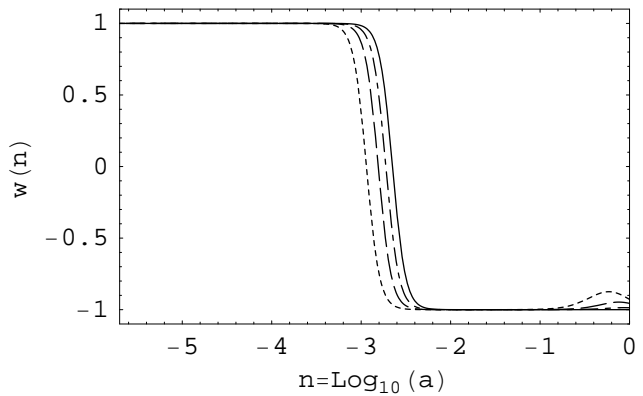}
\caption{We show the evolution of $w$ as a function of
$n=log_{10}(a)$ from $n_c$ (i.e. at $\Lmc$)
   to present day $n_o=0$
   for different initial conditions $\Omqi=0.01,\, 0.03,\,0.06, \,0.11$ (dotted, dashed,
   dot-dashed and solid    lines, respectively).  }
 \label{grsugraw}
\end{figure}
\begin{figure}[tbp]
    \includegraphics[width=8.5cm] {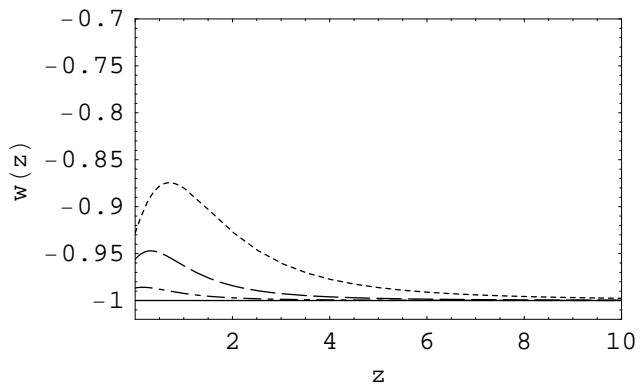}
 \caption{We show the late time evolution of $w$ as a function of redshift $z$
   for different initial conditions $\Omqi=0.01,\, 0.03,\,0.06, \,0.11$ (dotted, dashed,
   dot-dashed and solid    lines, respectively).  }
 \label{grsugrawz}
\end{figure}

In fig.\ref{grsugraw} we show the evolution of $w$ below $\Lmc$,
i.e. for $a> a_c$ (above $\Lmc$ it is simply $w=1/3$). It has an
initial kinetic period $w=1$ following a region with $w=-1$,
lasting almost the same amount of e-folds as the first one, and it
grows later to its present value $w_o$. We see that the value of
$w_o$ depends slightly on the initial conditions $\Omqi$. At
equivalence $ Log_{10}[a_{eq}]\simeq -3.6$ the value of dark
energy is  $\Omq\simeq 10^{-6}$ while at last scattering ($a\simeq
1/1090$) it is $\Omq\simeq 10^{-7}$, quite  small   in both cases.
In fig.\ref{grsugrawz} we show how $w$ increases as a function of
redshift $z$ for different values of $\Omqi $. In
fig.\ref{grsugrawdw}a we plot the present day values of $w_o$ and
$w'_o\equiv\pp w/\pp z|_o$  for different   initial $\Omqi$ and
final $\Omqo$. Notice that the variation is very small. In fact we
have $-1<w_o<-0.92$ and  $-0.06<w_o'<0.19$ if $\Omqo$ varies from
$0.65 \leq \Omqo \leq 0.75$ and $\Omqi<0.11$. For larger
$\Omqi>0.1$ we get a flat and  small  $w_o$, very close to -1,
with a slightly negative $w_o'$ while for   values   $\Omqi <0.1$
we   have a small positive $w_o'$ and a larger $w_o$. As seen from
fig.\ref{grsugrawdw}a the variation is more sensitive with respect
to $\Omqi$ than to $\Omqo$. The values of $w_o,\,w'_o$ are
insensitive to $H_o$. At this point, we would like to emphasis,
again, that  $\Omqi$ is not a free parameter, we do not
marginalize  it, but  we   show the slight dependence of
$w_o,\;w_o'$ over it.

We now compare our model with the data. The likelihood contours
for the Golden SN1a set \ci{sn1a} of $\Omqi\; vs\; \Omqo$ are
shown in fig.\ref{grsugraOfiOfo}. The best fit to the Golden SN1a
set for different initial conditions are shown in table
\ref{tab:sugSN}.  We see that for the different values of $\Ompi$
we get an equivalent fit as a LCDM. The acceleration
redshift (where the universe begins to accelerate) is centered
around $z_{ac}=0.65\pm 0.05$ with $\chi^2_{sn1a}<178$. The best fit
model has a very flat $w$ with $-1< w_o < -0.94$, even though
the $\Omqi=0.01$ case with $w_o=-9.94,\,w'_o=0.15$ does a reasonable job.

\begin{figure}[tbp]
    \includegraphics[width=8.5cm]{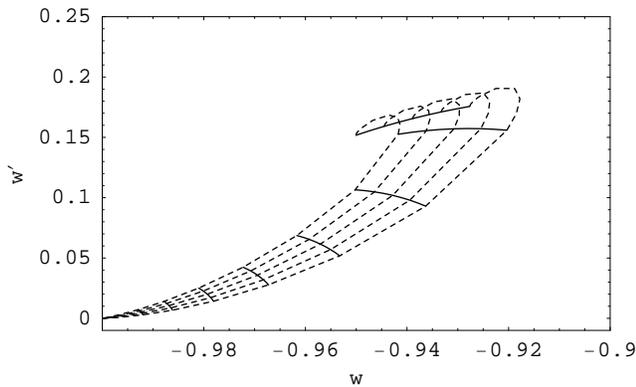}
  \caption{We  show   the values of $w$ and its derivative $w'=\pp w/\pp z|_o$ at present
  time
  for different values of $\Om_{Qo}$ and $ \Omqi$. The  dotted lines   have constant
  $\Om_{Qo}$ and from bottom to top they have $\Om_{Qo}=0.66,0.68,0.70,0.72,
  0.74$, respectively. The solid lines    have $\Omqi$ constant
  with $\Omqi=0.0002,\,0.01,0.02,0.03,0.04,0.05,0.06$ from top
  to bottom. Notice the resulting    $w,w'$ region is very small and it has
  an upper limit to $w$ and $w'$.  }
 \label{grsugrawdw}
\end{figure}
\begin{figure}[tbp]
      \includegraphics[width=8.5cm]{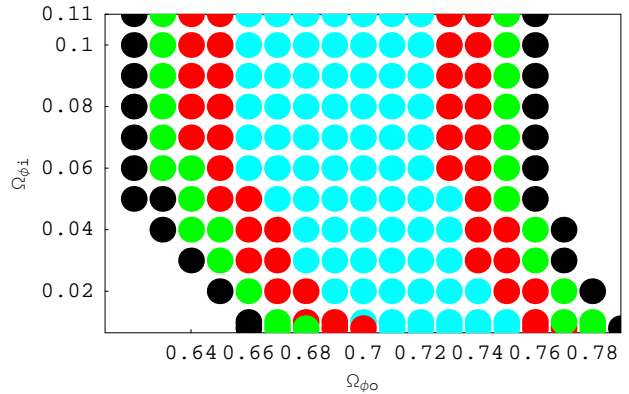}
  \caption{ We show the likelihood contours as a function of $\Omqi, \Om_{Qo}$
  compared to the Golden set of SN1a. We have
  $\chi^2_{sn1a}<178$,
  $\;178<\chi^2_{sn1a}<179$, $\;179<\chi^2_{sn1a}<180$ and $180<\chi^2_{sn1a}<181$
  (light blue, red, light green and black, respectively).}
 \label{grsugraOfiOfo}
\end{figure}
\begin{table}
 \begin{tabular}{|c|c|c|c|c|c|}
   \hline
    $\Omqi$ & $\Omqo$ & $\chi^2_{sn1a}$ & $w_o$ & $w'_o$ & $z_{ac}$ \\
   \hline\hline
     0.11 & 0.69 &177.1 & -1 & $-10^{-5}$ & 0.64  \\
     0.06 & 0.70 &177.2 & -0.99  & 0.01 & 0.67  \\
      0.03 & 0.71 &177.4 & -0.96  & 0.06 & 0.70  \\
     0.01 & 0.73 &177.5 & -0.94& 0.15 & 0.73  \\
   LCDM &   0.69 &177.1 & -1 & 0 & 0.65  \\
   \hline
\end{tabular}
\caption {  We show the best fit model compared to the Golden SN1a
set for  different values of $\Omqi$ and LCDM. } \label{tab:sugSN}
\end{table}
\begin{table}
\begin{tabular}{|c|c|c|c|c|c|c|c|c|}
  \hline
$\Omqi$ & $\Omqo$ & $\chi^2_{wmap}$ & $H_o  $ & $w_b  $ & $n_s  $
& $\tau $ &
$\sigma_8$& $w_o$ \\
 \hline \hline
  0.11 & 0.70 & 1428.7 & 69 & 0.0227 & 0.97 & 0.11 & 0.85 & -1 \\
  0.06 & 0.70 & 1428.7 & 69 & 0.0227 & 0.97 & 0.11 & 0.84 & -0.99  \\
  0.03 & 0.69 & 1428.7 & 68 & 0.023 & 0.97 & 0.11 & 0.83 & -0.96 \\
   0.01 & 0.71 & 1429.2 & 68 & 0.0227 & 0.97 & 0.13 & 0.77 & -0.93 \\
  LCDM & 0.70 &  1428.7 & 69 & 0.02327 & 0.97 & 0.11 & 0.84 & -1 \\
  \hline
\end{tabular}
\caption {  We show the best fit model compared to wmap data  for
 different values of $\Omqi$ and LCDM.  }
\label{tab:sugWmap}
\end{table}

To compare our model to the wmap data we use
  KINKFAST \ci{kink} (a modification of CMBFAST \ci{cmbfast} to include a dynamical
dark energy). The best fit to wmap  of our dark energy model and
LCDM can be seen from table \ref{tab:sugWmap}. The parameters  $n_s, \tau$ are degenerated
and different values would give the same $\chi_{tot}$ \ci{tegmark}.

\begin{table}
\begin{tabular}{|c|c|c|c|c|c|c|c|c|c|c|}
  \hline
$\Omqi$ & $\Omqo$ & $\chi^2_{wmap}$ & $\chi^2_{sn1a}$ &
$\chi^2_{tot}$ & $H_o  $   & $\tau $ &
$\sigma_8$ & $w_o$ \\
 \hline \hline
0.11 & 0.70 & 1428.7 & 177.1 & 1605.8  &  69 &   0.11 & 0.85 & -1 \\
0.06 & 0.70 & 1428.7 & 177.2 & 1605.9    & 69 &  0.11 & 0.84 & -0.99  \\
0.03 & 0.71 & 1428.8 & 177.4 & 1606.2  & 69 &   0.12 & 0.81 & -0.96 \\
0.01 & 0.72 & 1429.3 & 177.5 & 1606.8  & 69 &  0.12 & 0.76 & -0.94 \\
LCDM & 0.70 & 1428.7 & 177.1 & 1605.8 & 69  &   0.11 & 0.84 & -1 \\
  \hline
\end{tabular}
\caption {  We show the best fit model compared to the Golden SN1a set
and wmap
for   different values of $\Omqi$ and a LCDM. All models
have $n_s=0.97, \, w_b=0.0227$.
The number of degrees of freedom is the same for our dark energy model
and  LCDM, with a total dof 1342+157-6= 1493, giving a $\chi^2_{tot}/dof= 1.08$.}
\label{tab:sugTot}
\end{table}

Combining  the Golden SN1a set with wmap we show in table
\ref{tab:sugTot}  the best fit model  for   different values of
$\Omqi$ and for
 the best LCDM. Notice that all four initial conditions have an equivalent
fit and therefore, the data is not sensitive enough to distinguish
between them.
The total number of degrees of freedom
is $157$ for the Golden SN1a set and $1342$ for the wmap data. We
get for our best fit model a $\chi^2_{tot}=1605.8$ with $\chi^2_{tot}/dof= 1.08$.
The dof  is the same for our dark energy
model as  for LCDM. It is interesting to note that
after an exhaustive analysis of
model independent evolution of dark energy \ci{wz} the  best fit model
has $\chi^2_{tot}=1602.9$ at the price of introducing 4 extra parameters
giving the same  $\chi^2_{tot}/dof= 1.08$ but a worse
 Akaika or Bayesian criteria \ci{wz}. The Akaika information criteria \ci{aic}
 requires the smallest $AIC=\chi^2_{tot} + 2k$ while the Bayesion \ci{bay}
has $BIC=\chi^2_{tot} + k\,ln N$, where $k$ is the  number of paramters
and $N$ the number of data points. Our dark energy model and LCDM
have $AIC=1605.8+12=1617.8, \;BIC =1664.3$ while
the best $w(z)$ has $AIC= 1622.9, \;BIC= 1676$. We see that
for both criteria our dark energy models   has a better fit than
the best effective $w(z)$ model and an equivalent fit as  LCDM.
The evolution of our dark energy model and $\sigma_8$ lie at the central
values obtained in \ci{tegmark}-\ci{seljak}.

As a matter of completeness we have determined the best fit
if the exponential factor in eq.(\ref{vsugra}) is not present
(i.e. for $V=c^2\Lmc^{4+2/3}\phi^{-3/2}$). In this case we get
for $\Omqi=0.11$ a $P_5=(69,0.71,0.0228,0.12,0.97)$ giving $\chi^2_{tot}=\chi^2_{sn1a}+
\chi^2_{wmap}=177.7+1428.9=1606.6$ with $\chi^2_{tot}/dof=1.08$ and
  an equivalent fit as before.

To see the difference between our dark energy model compared to LCDM
we have calculated the CMB
using the same $P_6$ in both cases.
For small $\Omqi$ (i.e. $\Omqi=0.01,0.03$) we get a variation of
about $0.2-0.5\%$ increase in the three peaks while an increase of
about $1-3\%$ at lower multipoles  with a resulting $\Delta \chi^2_{wmap}\simeq 10$
in favor of our dark energy model. On the other hand for $\Omqi=0.06$ or $0.11 $
the variation is  much smaller, lees than $0.05\%$, giving, therefore, an
equivalent $\chi^2_{wmap}$. However, by changing $P_6$ we can get an
equivalent fit for LCDM as for our dark energy models.
The ISW effect for a dynamical $w(z)$ gives a nontrivial (but small)
imprint on the power spectrum \ci{wz} but this contribution
can be taking into account by  a change in $P_6$.
The great degeneracy on the cosmological parameters $P_6$
reduces the possibility of distinguishing between different dark energy
models at present time.  The  LSS data do not place a stronger constraint
on  the  evolution of dark energy  than  the SN1a and wmap data  \ci{tegmark}-\ci{seljak}.
Without an  independent way of measuring $P_6$ it is hard to distinguish
from the data between different dark energy models and we should
choose the best theoretical model.

Let us conclude. We have presented a realistic dark energy model
derived from particle physics
 that has essentially no free parameters. The reason why the universe
accelerates at such a late time is because the dark energy
(quintessence field) appears at a very late time due to the gauge
group dynamics. Our model has a almost flat $w$, with $-1<w_o
<-0.94$, and has a better fit to the cosmological observations
than the best $w(z)$ effective model and  an equivalent fit than
LCDM. Unfortunately the great degeneracy on the cosmological
parameters $P_6$ reduces the possibility of distinguishing between
different dark energy models from the observations.  Therefore, we
are left, at present time, with the most appealing   theoretical
model as the best candidate for  dark energy. This is the model
with the least number of free parameters and with the best
explanation of the origin of dark energy and its late time
appearance. Our dark energy model satisfies these criteria.

\begin{acknowledgments}
We would like to thank P.S. Corasaniti for making
KINKFAST1.0.1 available  and  for useful discussions. This work was supported in part by CONACYT project 45178 and
DGAPA, UNAM project IN-110200.
\end{acknowledgments}

\end{document}